\documentclass[onecolumn]{aa}
\newcommand{\kms}{km s$^{-1}\;$}
\newcommand{\kmss}{km s$^{-1}$}
\newcommand{\km}{km s$^{-1}\;$}

\newcommand{\vlsr}{V$_{\rm LSR}$}

\newcommand{\lsun}{\mbox{L$_{\sun}$}}

\newcommand{\ho}{H$_{2}$O$\;$}

\usepackage{graphicx}
\begin{document}
  \title{Location of \ho\ maser in the double-nuclei system  of NGC\,6240}
   \author{Y. Hagiwara
          \inst{1},
%          \and
          P. J. Diamond 
          \inst{2},
          \and
          M. Miyoshi
          \inst{3}
          }

   \offprints{Y. Hagiwara}

   \institute{ASTRON, Westerbork Observatory, P.O. Box 2, 7990 AA
         Dwingeloo, The Netherlands \email{hagiwara@astron.nl} 
         \and      
       Jodrell Bank Observatory, University of Manchester,
        Macclesfield, Cheshire, SK11 9DL, UK
         \and  National Astronomical Observatory,
             2-21-1 Osawa, Mitaka, Tokyo, Japan,
             181-8588}
%   \date{}
\authorrunning{Y. Hagiwara et al.}
\titlerunning{Water maser in the double nuclei of  NGC\,6240}

\abstract{We performed VLA observations of the 22 GHz \ho maser emission in
   the merging galaxy NGC\,6240, which hosts the well-known double active nuclei. In a
   previous paper, we reported on the first solid detection of the \ho
   maser emission in 2001. After two
   abortive attempts due to the weakness and probable variability of
   the emission, the maser was detected with the VLA in June 2002. The
   emission is unresolved at $\sim$ 0.3 arcsecond and 
   coincides with the southern 22 GHz continuum peak to $\sim$ 0.007
   arcsecond ($\sim$ 3 pc: D = 97 Mpc).  The detection of the maser in
   the southern nucleus indicates that nuclear activity of the galaxy,
   which is significant in X-ray and far-infrared (FIR) bands, lies
   mainly in the southern nucleus, and the nucleus without a high
   brightness peak could be explained by thick dust emitting FIR
   radiation.  We favour the idea that the maser in NGC\,6240 is 
   associated with the AGN-activity.
 }

   \maketitle
%
%________________________________________________________________
%
%
\section{Introduction}
\subsection{Background of extragalactic \ho masers}
A number of attempts to find more extragalactic \ho masers (6$_{16}$--5$_{23}$) have been made in recent times,
stimulated by the discovery of the highly Doppler-shifted maser
emission symmetrically straddling the systemic velocity of the galaxy
in NGC\,4258 (\cite{naka93}). VLBI observations of the maser emission
in NGC\,4258 revealed the presence of a sub-parsec-scale \ho maser
disk rotating around a central massive object (\cite{miyo95}). After
the discovery of the masing torus surrounding a 'black hole' in
NGC\,4258, very few other such sources suitable for studying the
central sub-parsecs of Active Galactic Nuclei (AGN) have been
discovered, but astronomers need more candidates to probe the full
range of the kinematics and dynamical structures of sub-parsec-scale
circumnuclear regions in AGN. This kind of study is possible through
the technique of radio interferometry. The extragalactic \ho masers observed and studied to date can be categorized into two types in terms
of the origin of the emission. Most of $low-luminosity$ masers (L$_{\rm {H_2O}}$ $<$ $\sim$ 1 \lsun) have been observed outside galactic nuclear regions, and are associated with star forming regions and (compact-)HII regions, suggesting that the masers have no direct connection to AGN-activity. $High-luminosity$ masers (L$_{\rm {H_2O}}$ $>>$ 10 \lsun) are apt to be located in or nearby an active nucleus, or to be associated with jet-activity. Such a water maser which associates with the AGN-activity is a $nuclear \; maser$. In either case they trace the nuclear activity within the central few parsecs of the AGN. There are also a number of masers with luminosity on the order of 10 \lsun, for instance, those in Mrk\,1 and NGC\,5506 (\cite{braa94}).

The origins of \ho masers with luminosities in the range 10 L$_{\odot}$ $<$
L$_{\rm {H_2O}}$ $<$ 100 L$_{\odot}$ have not been well studied in comparison with those of higher luminosity because the \ho emission is generally not
sufficiently bright for radio interferometric observations with
reasonable sensitivities. Intensity variabilities on time scales of
weeks to months are commonly observed in these masers, which has posed
difficulties in determining the optimum timing of interferometric
experiments.  Currently, about 25--30 extragalactic \ho masers,
excluding the low luminosity masers, are known. However, only about
ten of them have been so far imaged at milliarcsecond resolution using
VLBI (\cite{mora99,linc00,hagi01,ishi01}). Radio interferometric
studies of this sort of \ho maser are quite challenging, but are
necessary to investigate the overall properties of the \ho masers in
active galaxies regardless of their maser intensity.  The apparent
luminosity of the maser is not always relevant to its intrinsic
properties.\\
In a previous paper, we reported on the single-dish detection of the \ho
maser in NGC\,6240 (\cite{hagi02}), and here present the
results of follow-up observations at subarcsecond angular resolution
aimed at investigating the spatial distribution of the maser.
\subsection{NGC\,6240}
NGC\,6240 (D=97 Mpc, assuming that H$_0$ is 75 \kmss Mpc$^{-1}$;
0.1''=48.5~pc) is a well-studied galaxy at various wavelengths. It is
a prototypical ultraluminous FIR galaxy with a complex optical structure, hosting a LINER nucleus (\cite{heck87}). The huge FIR luminosity ($> 10^{11}$ \lsun) is considered to originate from
re-emission of ultraviolet radiation processed by a dust shell
surrounding a hot core, suggesting ongoing star formation in the
galaxy (e.g., \cite{sand88}). X-ray satellite observations detected 6.4
keV Fe K$\alpha$ line emission in the galaxy, showing a definite
presence of an AGN together with a large X-ray luminosity of $\sim$ 10$^{43}$ - 10$^{44}$ erg s$^{-1}$ in the 2 - 10 keV bands (e.g., \cite{iwas98}). At sub-arcsecond resolution a number of hard X-ray blobs were imaged by Chandra, and some of them extend several kiloparsec from the center of the galaxy (\cite{lira02}).  5~GHz high-resolution continuum MERLIN maps made by Beswick et al. (2001) showed two
compact radio sources (double nuclei) with non-flat spectral indices ($\sim$ 0.7 -- 0.8 at 4.9 --15 GHz (\cite{colb94}, \cite{carr90})), and they are separated by 1''.5 at the highest resolution so far attained ($\sim$ 0.1 arcsecond)
in the nuclear region. The velocity fields of the broad HI absorption
towards these components are seen running along the position angle
joining the two radio nuclei. Beswick et al. (2001) found that the
radio continuum flux in the galaxy is mostly dominated by a starburst
component, with only a small portion arising in a low luminosity
AGN. Tacconi et al. (1999) and Bryant et al. (1999) found a
significant peak of CO molecular gas between the double radio
nuclei. Their speculation is that the CO gas settling down
between the two stellar nuclei will eventually form a central thin
disk. With lower angular resolutions but higher sensitivities than
MERLIN, the VLA revealed HI and OH absorption structures, which agree
with those of H$_2$, CO, and H$_{\alpha}$ (\cite{baan02}). There has
been an open discussion on the need to introduce an AGN in NGC\,6240
as an energy supplier sustaining the large infrared luminosity,
however, no one has yet succeeded in determining the location of the
predicted AGN in the double nuclei system at any wavelengths. 
\subsection{\ho maser in NGC\,6240}
Tentative detections of an \ho maser were made by Henkel et al. (1984)
and Braatz et al. (1994), although never confirmed. Hagiwara et
al. (2002a) reported on the first clear detection of the water maser in
the galaxy in May 2002. They also detected intensity variability on
timescales of weeks to months by means of single-dish monitoring of
the two distinct components, one of which lies at \vlsr = 7565 \kms
and the other at \vlsr = 7609 \kmss. Given the fact that the apparent
maser luminosity of the two components is about 40 \lsun, well above that of the typical starforming masers, and the emission is highly redshifted by some 400
\kms with respect to the adopted systemic velocity of \vlsr = 7131
\kms (\cite{hagi02}), the maser is likely to arise from the AGN activity and can be used to probe the kinematics in the galaxy.
\section{Observations and data analysis}
\subsection{The 100-m radio telescope in Effelsberg}
After the first single-dish detection of the \ho maser emission in the rotational 6$_{16}$ -- 5$_{23}$ transition (rest frequency = 22.23508 GHz) towards NGC\,6240 in May 2001 (\cite{hagi02}), we observed the maser at several epochs until September 2002. All the observations were performed with the Max-Planck-Institut f\"ur Radio astronomie (MPIfR) 100-m radio telescope in Effelsberg using a two channel K-band HEMT receiver with dual polarizations and a digital autocorrelator (AK90) as a backend. The backend correlator provided a total bandwidth of 4 $\times$ 40 MHz for each polarization (velocity coverage $\simeq$ 2100 \kmss) with a spectral resolution of 78 kHz (1.1 \kmss). The beam size of the telescope is so small ($\sim$ 40\arcsec (HPBW) at 22 GHz) that the frequent calibration of telescope pointing was critical. It was determined by observing a nearby quasar 1655+077 every 1.5 -- 2 hrs during the observations, yielding the pointing accuracy of 4\arcsec -- 6\arcsec. More details of the observing parameters were remarked in Hagiwara et al. (2002a). Data analysis consisted of averaging the spectra and subtracting a linear baseline in general, but occasionally a non-linear baseline.
\subsection{Very Large Array}
NRAO\footnote{The National Radio Astronomy Observatory (NRAO) is
operated by Associated Universities, Inc., under a cooperative
agreement with the National Science Foundation.}  Very Large Array (VLA) observations
were carried out in the B configuration on 16 July 2002.  Prior to
that epoch, VLA or VLBA observations had been attempted several times
in 2001 with no clear detections; a list of observations is displayed
in Table 1.  The current observations (this paper) employed a single
IF band of 12.5 MHz divided into 64 channels of width 2.64 \kms\ each.
The band was centered at an LSR velocity (\vlsr) of 7585 \kmss, which
is about 450 \kms redshifted from the systemic velocity of the
galaxy. After removal of several channels at the band edges, the
usable velocity range was 7510 -- 7665 \kmss.  To detect maser lines
with weak flux densities of $<$ 50 mJy, we conducted phase-referencing
observations using a nearby quasar, 16582+074 as a phase-reference
source. The phase-referencing observations were executed in a sequence
of 3 min scans with cycling interval of 140 sec for NGC\,6240 and 40
sec for 16582+074. The total observing time was three hours, and the
resultant synthesized beam has a FWHM size of
$0\rlap{.}{''}29\times0\rlap{.}{''}26$ with uniform weighting and a
position angle of $-$34${^\circ}$. The amplitude and bandpass
calibrations were made by observing 3C\,286, assuming $S({\rm
3C\,286}) =$ 2.59 Jy at 21.7 GHz.  The data were calibrated and mapped
in the standard way using AIPS. The continuum emission was subtracted
from the spectral line visibility cube using UVLSF, and then velocity
channel maps of the H$_2$O emission were made for each velocity
channel.  The resultant 64 channel maps each had a typical rms noise
level of 1.3 mJy beam$^{-1}$. The continuum emission was imaged with
an rms noise of 0.26 mJy beam$^{-1}$.
\section{Results}
\subsection{Effelsberg}
  Fig. 1 shows \ho maser spectra obtained from the MPIfR
100-m telescope for 4 epochs between 10 May 2001 and 14 September 2002. Each spectrum contains the two distinct redshifted components at \vlsr = 7565 \km and 7609 \km and other weaker components that require confirmation. We searched for maser emission primary from \vlsr $\simeq$ 6800 -- 7800 \kmss. Occasionally, the highly blueshifted velocities ranging to \vlsr = 5900 \kms were searched (Table 1). No emission was distinctly detected except for the two  components with 3 $\sigma$ limits of 15 -- 60 mJy. The spectra of the two epochs in 2001 have better signal-to-noise ratios than those in 2002 largely because the integration times are much longer. Given the fact that the velocities of the two components were not changed over 8 months, an upper limit of drift rate of each line is 1.3 \kmss ~year$^{-1}$. The flux density of the \vlsr = 7565 \km component was measured to be $\sim$ 50 mJy in March 2002, the highest value recorded during our single-dish monitoring. The maser was not detected in September 2002 with an rms of $\sim$ 20 mJy. Neither component was detected by the VLA snapshot made in early Oct 2001 with a 3 $\sigma$ level $\sim$ 15 mJy per 2.6 \kms spectral channel. This implies that the maser once faded after July 17 2001.
\subsection{VLA}
A naturally weighted VLA 22~GHz continuum map is shown in Fig. 2 at
0.3'' resolution; it displays double-peaked radio emission in the
nuclear region in radio robes around each peak. The radio morphology is consistent with the 8.4 GHz and 15~GHz continuum images observed with
the VLA in A configuration (\cite{colb94}).  The components 'S' and
'N' in Fig.2 correspond to N1 and N2 in Fig.2 or Fig. 3 of Colbert et
al. (1994). Spectra of the 22 GHz maser emission towards S and N are
also shown in Fig.2, both of which were determined from the areas
containing the continuum peaks. 

The maser line spectrum towards S in Fig.2 shows a single line profile with a flux density of about 35 mJy (L$_{\rm {H_2O}}$ $\simeq$ 20 \lsun) , and a centroid velocity of \vlsr = 7611 \kmss, while
the spectrum towards N does not show any emission within a 3$\sigma$
level of $\sim$ 4 mJy per velocity channel (L$_{\rm {H_2O}}$  $\la$ 2 \lsun), suggesting that N houses no high luminosity maser at this time.  Since the velocity resolutions of the 100-m and VLA spectra are different, there is the ambiguity in estimating the centroid velocity of each line in each spectrum. The uncertainties of velocity are not better than the VLA channel spacings of 2.64 \kmss.  Therefore, we conclude that the VLA component lying at \vlsr = 7611 \kms is identical to the 7609 \kms component detected by the 100-m and shown in Fig.1.  The position of the maser is $(\alpha,\delta)_{\rm J2000}$ = $16^{\rm h}52^{\rm m}58\rlap{.}^{\rm s}88 ~\pm$ 0.$^{\rm s}$05,
$+02^\circ24'03\rlap{.}\arcsec 3 \pm 0\arcsec.1$. The relative positional error, dominated by statistical noise, between the maser and the 22 GHz
continuum peak S is estimated to be $\sim$ 0.007 arcsec.  Accordingly,
the maser coincides with the continuum peak within an astrometric
uncertainty of $\sim$ 0.007 arcsec, or $\sim$ 3 pc. The maser emission
remains unresolved at our angular resolution of 0.26'', or 126
pc. Assuming a Gaussian distribution of the \ho emission, the source
size deconvolved by the clean beam is 0.12 $\times$ 0.08 arcsec (58 pc
$\times$ 39 pc). \ho line emission was not detected with a 4 $\sigma$
limit of 5.2 mJy from any other observed points in the galaxy,
including towards N. We reported earlier on the marginal
interferometric detection of the \vlsr = 7565 \km component obtained
by the VLA (C configuration) in Aug 2001, the position of which was
significantly displaced from both S and N (\cite{hagi02b}). The
sensitivity of that map is, however, quite poor, and the detection
level of the emission was $\sim$ 3 $\sigma$. We will not use that
result in the later discussion.

Time variability of the maser intensity is that the two
distinct components show changes in flux density from 10 -- 50 mJy
over weeks to months. There seems to be a weak anti-correlation of
intensities between the two components.
\section{Discussion}
\subsection{Location of the \ho maser in NGC\,6240}
The most significant result from our high-resolution imaging of
NGC\,6240 is that the spatially unresolved maser is exactly coincident
with the center of the southern nucleus (S) and there is no detectable
\ho emission from the northern nucleus (N). There is compelling
evidence that luminous \ho\ maser emission marks the radio continuum nucleus of the galaxy as in NGC\,4258 (e.g., \cite{miyo95}). The detected maser in
 NGC\,6240  may well probe the position of one of the  double-nuclei, S. Given
the detection of features redshifted by $<$ 400 \km from the
systemic velocity and the narrowness of each line, the maser might
originate from a well-defined region, for example, a tangential point
in the receding side of a rotating disk. In addition such extremely high velocity shifts of each line  have never been measured in stellar jets or outflows. Alternatively, it is possible
that the redshifted maser emission is associated with the jet-activity
of AGN. We cannot support the latter at this moment because there has not been detection of any spatially resolved core-jet structure in S and N at the highest resolution of 0.055 arcsec (\cite{besw01}). The interferometric observing results at different radio frequencies leave the possibility that these two radio sources are stellar, i.e. non-galactic, components as they were resolved by milliarcsecond VLBI observations at 1.4 GHz resulting in non-detection of ultra-compact radio sources on scales of 0.5 pc -- 5 pc in N and S
(unpublished results; Hagiwara et al.), and neither N nor S shows the
flat radio spectral indices (Table 2) that are expected for such
resolved nuclei (\cite{colb94}). The apparent \ho maser luminosity of 20 \lsun\ is nevertheless brighter by two orders of magnitude than
that of starforming masers such as NGC\,253 and M\,82; their typical
\ho maser luminosity is $\sim$ 0.1 \lsun\ or less, and neither of them
lies in a nucleus, i.e. the dynamical center of the galaxy
(\cite{ho87,baud96}).  If S is one of the real nuclei of the galaxy, the
interpretation of the maser will be straightforward - a nuclear maser 
with the emission physically confined within 1 pc of S. It is plausible that the other maser component will be detected towards N. The absence of compact radio emission between N and S is not consistent with the detection of the CO (1-0) between them, but a radio-quiet AGN is not ruled out.
The dust shell
foreground to the double peak system which is responsible for the
strong far-infrared emission may hide a true nucleus in the
background, or obscure these nuclei. The dusty environment itself is
favorable to maser emission, and the maser could be enhanced through
the thick dust lane in our line of sight, amplifying the background
continuum source. It is interesting that the dust in the galaxy causes
highly variable extinction with A$_{V}$ = 2 -- 8 at optical bands 
(\cite{scov00}).
\subsection{Physical conditions for the luminous \ho\ maser}
The observation that the presence of extragalactic OH and \ho maser
emission is mutually exclusive appears to be due to the different
properties of host galaxies: the OH megamaser galaxies show large FIR
luminosities and gas-rich properties in the circumnuclear region but
do not harbor a distinct active nucleus, while the \ho maser galaxies
on the whole contain an AGN which provides the conditions necessary
for maser amplification. NGC\,6240 however exhibits both OH and \ho
molecular gas towards the double nuclei system. OH molecules in the
galaxy are not seen to be producing maser emission but exist as
absorbing gas that is extended on kiloparsec scales, suggesting that
the water masing cloud and OH absorbers do not coexist.  In general,
do OH megamasers coexist with \ho megamasers under similar physical
environments? We note one theoretical study by Neufeld et al. (1994)
on the dense circumnuclear gas in a thin gaseous layer, illuminated by
X-rays from AGN, which produces the conditions that are necessary for
\ho maser excitation. When the temperature of a molecular gas phase
heated by nuclear X-ray radiation from a central engine rises
sufficiently ($\sim$ 400 -- 1000 K), a pair of chemical reactions of O
$+$ H$_{\rm 2}$ = OH $+$ H and OH $+$ H$_{\rm 2}$ = \ho $+$ H will
occur (e.g., \cite{elit92}). The density of OH, hence, will be
relatively smaller, and the abundance ratio of water to hydrogen will
be greater than 10$^{-4}$ as the reaction progresses. Note that the
temperature required to enhance the reactions is $>$ 400 K, but under
such conditions the OH molecule is less likely to emit as a
maser. With such \ho abundance, the apparent luminosity of the
saturated \ho maser can range from about 30 - 300 \lsun\ pc$^{-2}$
over a wide range of parameters of X-ray fluxes, gas pressures,
densities, gas temperatures, and effective scales of X-ray heating
from the surface of a slab of material, all of which are derived from
reasonable observational values. Even with the introduction of a torus
model instead of the slab, the luminosity will be of similar value
(\cite{wall97}). Such X-ray heating models producing luminous \ho
masers will not give rise to OH masers in the warm gas. \\  The high-luminosity maser must be located in a smaller area than the VLA synthesized beam (e.g., \cite{clau86}). The location of the unresolved maser is well
defined within a radius of the beam-deconvolved size of $\sim$ 30
pc.  If we adopt the total maser luminosity of 20 \lsun, the luminosity
per square parsec is L$_{\rm H_2O}$ $\la$ 0.01 \lsun\
pc$^{-2}$.  The relative location of the maser in NGC\,6240 is measured to be within 3 parsecs of its continuum peak, yielding L$_{\rm H_2O}$  $\simeq$ 0.7 \lsun\ pc$^{-2}$. Recent milliarcsecond VLBI studies have demonstrated that the distribution of \ho megamaser emission is, in general, confined to $\sim 1$ pc of the center of galaxies (\cite{mora99}). With this in mind, the lower limit of the NGC\,6240 maser luminosity will be
comparable to 20 \lsun\ pc$^{-2}$, which agrees with the minimum
value of the theoretically estimated luminosity. The \ho maser in
NGC\,6240 can, therefore, be explained by the AGN-activity.
\subsection{Gas kinematics in NGC\,6240}
If the maser is a nuclear maser, it should lie in the dynamical center
of the galaxy, where the observed dense molecular gas will eventually
settle down between the two radio sources and form a disk
(\cite{brya99,tacc99}). However, our data show that the maser lies in
one of the nuclei displaced by some 100 pc from the CO emission peak
lying between the twin nuclei. Is this the real dynamical center of
the galaxy?  It is understood that there is no well-identified active
nucleus in the galaxy, and the position of the detected 6.4 keV Fe
line emission is uncertain (\cite{ikeb00}). Beswick et al. (2001)
found different centroid velocities of the HI absorption towards N and
S; they peaked at \vlsr = 7260 \kms and 7087 \kmss, respectively. The
velocity ranges of the absorption do not overlap with any of the known
maser components.  The radial HI velocity towards S is blueshifted by  about 50 \kms from the galaxy's systemic velocity (\vlsr=7131
\kmss), while the velocity towards N is redshifted by about 130 \kms
to the systemic velocity (Table 2). The nucleus S clearly shows
different kinematics from N. This situation is similar to the radial
velocity difference  between the nuclei of 150 \kms in H$_2$
$\nu$ = 10 S(1) and 100 \kms in the CO (J = 2 $-$ 1) (\cite{tecz00}).

If the maser were to exhibit a symmetric velocity distribution, we
might assume that the components we detect at \vlsr = 7565 \kms and
7611 \kmss are red-shifted components and we might therefore search
for emission around the 'systemic' velocity of S (\vlsr $\approx$ 7080 \kmss),
and the blueshifted counterparts of the red-shifted emission around
\vlsr = 6500 \kms and 6550 \kmss. However, in the latest Effelsberg
observations of the maser  carried out on 14 September 2002, no new components were detected within $\pm$ 800 \kms centered on the systemic velocity to an rms of $\sim$ 20 mJy. On the other
hand, the velocities of CO emission (J = 1 $-$ 0, 2 $-$ 1) span a
velocity width of $\pm$ 500 \km centered on the systemic velocity,
reaching to the redshifted maser velocity.  The CO emission that peaks
between the twin nuclei seems to trace a circumnuclear region on
larger galactic disk scales. The orientation and position of the
galactic scale disks such as those imaged by HI, CO or HCN gas have
little correlation in general with those of a maser disk on sub-parsec
scales (e.g., \cite{hagi01b}). If the two radio nuclei are highly obscured by the thick dust lane and scattered by the foreground
materials in the galaxy, they will be observed as less intense due to
absorption and are likely to be resolved with milliarcsecond
resolution.  Note that the radio spectral properties of N and S are not significantly different (\cite{colb94}), suggesting that both of the nuclei could house AGN. 

In conclusion, we propose that one of the nuclei lies in S, and the AGN activity which is more dominant at S than any other locations in the galaxy gives rise to the maser emission. However, the starburst or composite starburst plus AGN components at the nuclei cannot be ruled out, despite the fact that such a luminous ($>$ 10 \lsun) water maser has not ever been observed in starburst galaxies (e.g., \cite{clau86}). It is essential
to measure the locations of all visible maser components with
VLBI. Our interpretation would have to be reconstructed if any of the
other velocity components are found to be located outside the nuclei in future
observations.
\begin{acknowledgements}
 YH thanks R. Beswick and W. Baan for useful comments and discussion. The authors would like to thank the referee, Jim Braatz, for his critical and careful reading of the manuscript.
\end{acknowledgements}
\onecolumn

\newpage
   \begin{figure}
%   \centering
   \includegraphics[angle=-90,width=16cm]{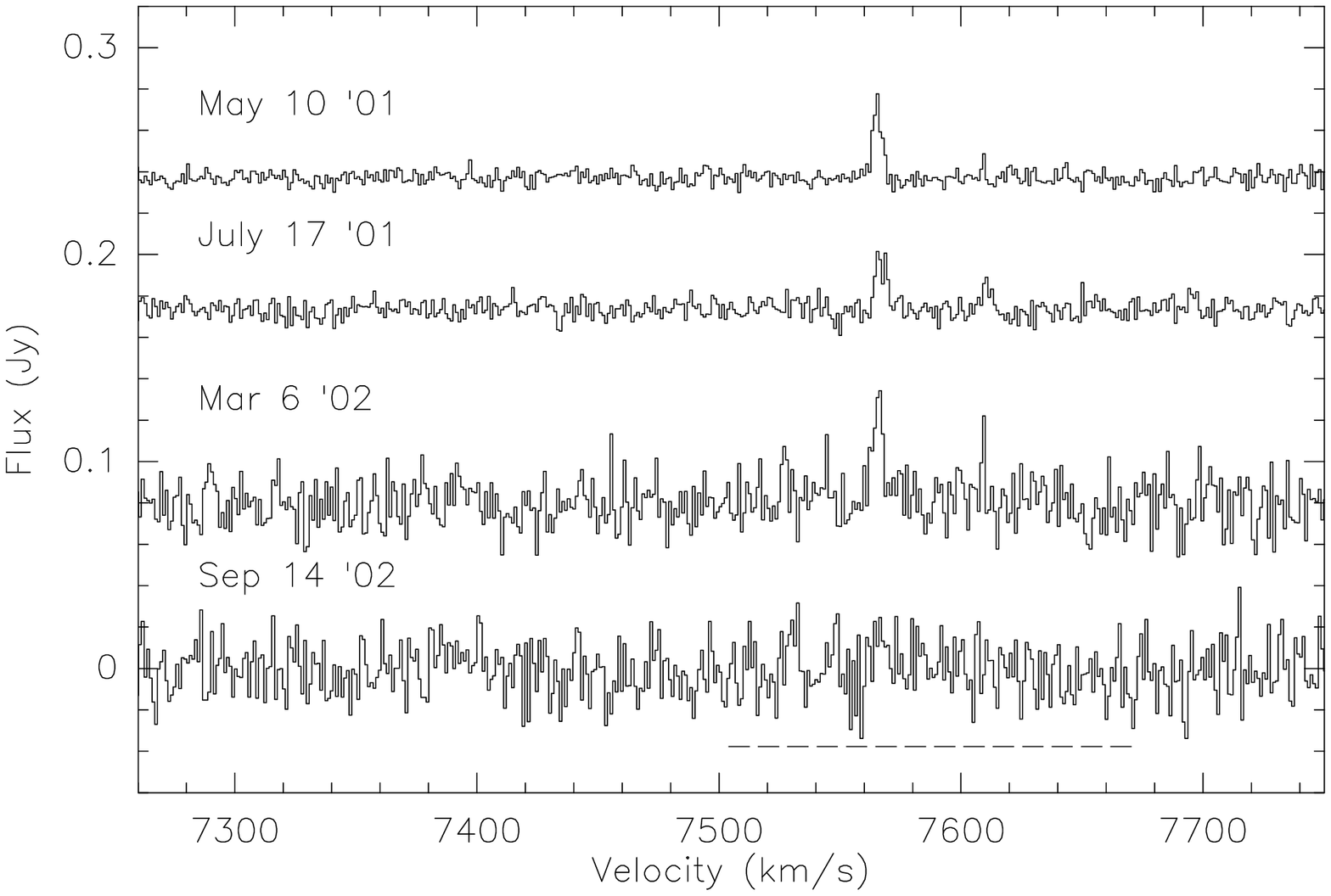}
      \caption{Spectra of all the detected maser components for 4 epochs of the single-dish observations. The systemic velocity of the galaxy is \vlsr=7131 \kmss in the radio definition. The spectral channel resolutions are 1.1 \kmss. The dotted line indicates the observed velocity range by the VLA. Two features of \vlsr = 7565 \kms and 7609 \kms were visible except at the fourth epoch. The VLA detected only the 7609 \kms feature between epochs March 6 and September 14 in 2002. The differences in the noise levels are primarily due to differences in the integration times.}         
         \label{fig1}
\end{figure}
   \begin{figure}
%
%   \centering
   \includegraphics[angle=0,width=15cm]{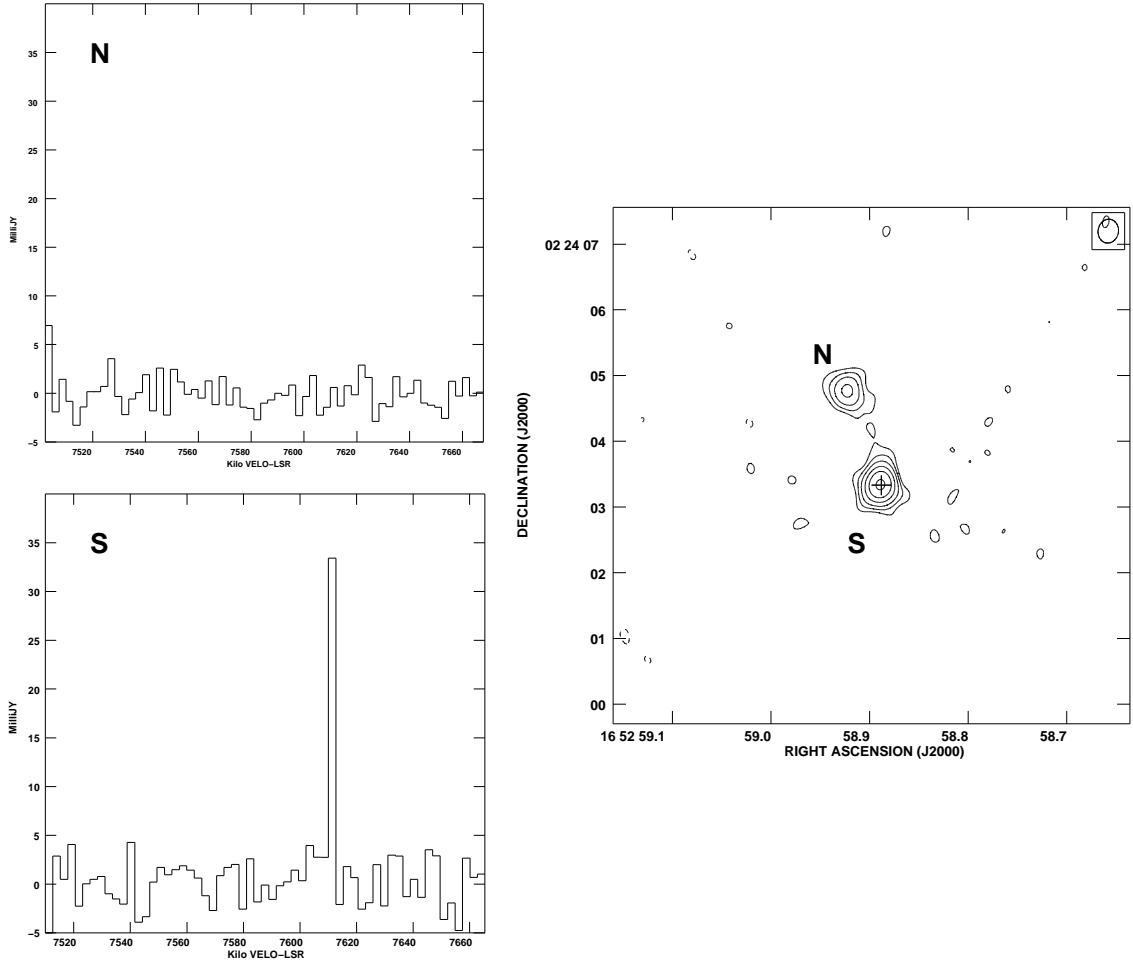}
      \caption{22 GHz radio continuum image of NGC\,6240. The synthesized beam is plotted in the upper right corner. Two \ho maser spectra are displayed against the double nuclei. The data were taken on 16 June 2002 by the VLA
      in B configuration. The velocity resolutions of the spectra are
      2.64 \kmss. The maser emission shows only one component
      centered on 7611 \kmss, which is probably identical to the one at 7609 \kms in Fig.1. Contours are -5.7, 5.7, 10, 17, 30, 51, and 90 \% of peak
      flux density of 12.3 mJy beam$^{-1}$. The position of the maser
      emission peak, marked by a cross in the image (The extent of the cross does not reflect position errors.), coincides exactly the southern continuum nucleus.}
         \label{fig2}
   \end{figure}
 \begin{table}
      \caption[]{Summary of observations}
         \label{t1}
\begin{tabular}{llllll}
            \hline
           \noalign{\smallskip}
            Telescope & Date & Configuration & RMS$^{\mathrm{a}}$ (mJy {beam}$^{-1}$)
           & Velocity range (\vlsr) & Comments \\
%            \noalign{\smallskip}
             \hline
%     \noalign{\smallskip}

Effelsberg  & 10~May~2001  &  & 5-7 & 6850 -- 7870 &  Hagiwara et al. (2002a)\\
Effelsberg  & 17~Jul~2001  &  & 5-7 & 6850 -- 7870 & Hagiwara et al. (2002a) \\
   VLA   & 5~Aug~2001  & C & $\sim$  5 &  7545 -- 7625 & 3 $\sigma$$^{\mathrm{b}}$ \\
   VLA   & 1~Oct~2001  & D &$\sim$ 5 &  7545 -- 7625 & Non-detection \\
    VLBA    & 4~Oct~2001  & VLBA 10 & $\sim$ 6 & 7255 -- 7355 &
   Non-detection    \\
     &   &  &  &  7535 -- 7635&
   Non-detection    \\
Effelsberg  & 4-6~Mar~2002  &  & $\sim$ 15 & 6700 -- 8000 &  \\
VLA   & 16~Jun~2002  & B & 1.3$^{\mathrm{c}}$ & 7525 -- 7665 & This paper \\
Effelsberg  & 14~Sep~2002  &  & $\sim$ 20 & 5900 -- 7750 & No components were detected \\
            \hline
\end{tabular}
\begin{list}{}{}

\item[$^{\rm{a}}$]{Rms noise level per velocity channel of 1.1 \kms (Effelsberg),  1.3 \kms (VLA), and 1.7 \kms (VLBA)  }
\item[$^{\rm{b}}$]{Detection level is insufficient for study in our research}
\item[$^{\rm{c}}$]{2.6 \kms spectral resolution}
%\item[$^{\rm{d}}$]{1.1 \kms spectral resolution}

\end{list}

\end{table}

\begin{table}
\caption[]{Double-nuclei in NGC\,6240}
\label{tbl2}
\[
  \begin{array}{p{0.5\linewidth}ll}
%            \hline

                             & $Northern~nucleus$(N)& $Southern~nucleus$(S)  \\

\hline
R.A. (J2000)$^{\mathrm{a}}$ & 14^h 52^{\rm m} 58 \fs 92 & 14^h 52^{\rm m} 58 \fs 89          \\
Dec. (J2000)$^{\mathrm{a}}$ & +02 \degr 24\arcmin 04.8 \arcsec & +02
\degr 24\arcmin 03.3 \arcsec \\
Systemic Velocity (21cm HI)$^{\mathrm{b}}$& 7260~$km$~$s$^{-1} & 7087~$km$~$s$^{-1}\\
Flux density (22 GHz continuum)$^{\mathrm{c}}$ &~~ 3.9~$mJy$ & 11~$mJy$  \\
Flux density (\ho maser) &~~~~ $--$  & 35~$mJy$  \\
Spectral index (4.9,8.4,15 GHz) $^{\mathrm{d}}$   &~~ 0.8 \pm 0.15 &
0.7 \pm 0.15 \\

\hline

\end{array}
\]
%\end{table}
\begin{list}{}{}
\item[$^{\mathrm{a}}$] Determined by VLA-B at 22~GHz (this paper)
\item[$^{\mathrm{b}}$] LSR velocity, assuming the radio definition (Beswick et al. 2001).
\item[$^{\mathrm{c}}$] Measured in a map with uniform weighting (this paper)
\item[$^{\mathrm{d}}$] Colbert et al. (1994)
\end{list}
\end{table}

\end{document}